# J3DAI: A tiny DNN-Based Edge AI Accelerator for 3D-Stacked CMOS Image Sensor

Benoit Tain[1], Raphaël Millet[1], Romain Lemaire[2], Michal Szczepanski[1], Laurent Alacoque[2], Emmanuel Pluchart[2], Sylvain Choisnet[2], Rohit Prasad[1], Jérôme Chossat[3], Pascal Pierunek[3], Pascal Vivet[2], Sébastien Thuries[2]
[1]Univ. Paris-Saclay, CEA, List, Palaiseau, France; [2]Univ. Grenoble Alpes, CEA, List, Grenoble, France;
[3]STMicroelectronics, Crolles, France; Email: benoit.tain@cea.fr

*Abstract*—**This paper presents J3DAI, a tiny deep neural network-based hardware accelerator for a 3-layer 3D-stacked CMOS image sensor featuring an artificial intelligence (AI) chip integrating a Deep Neural Network (DNN)-based accelerator. The DNN accelerator is designed to efficiently perform neural network tasks such as image classification and segmentation. This paper focuses on the digital system of J3DAI, highlighting its Performance-Power-Area (PPA) characteristics and showcasing advanced edge AI capabilities on a CMOS image sensor.**

**To support hardware, we utilized the Aidge comprehensive software framework, which enables the programming of both the host processor and the DNN accelerator. Aidge supports post-training quantization, significantly reducing memory footprint and computational complexity, making it crucial for deploying models on resource-constrained hardware like J3DAI.**

**Our experimental results demonstrate the versatility and efficiency of this innovative design in the field of edge AI, showcasing its potential to handle both simple and computationally intensive tasks.**

**Future work will focus on further optimizing the architecture and exploring new applications to fully leverage the capabilities of J3DAI. As edge AI continues to grow in importance, innovations like J3DAI will play a crucial role in enabling real-time, low-latency, and energy-efficient AI processing at the edge.**

## I. Introduction

The increasing adoption of intelligent vision systems in various domains, including the Internet of Things (IoT), healthcare, automotive applications, and smart surveillance, has led to an increasing demand for advanced image sensor technologies capable of real-time processing. Traditional cloud-based processing architectures face with several challenges, such as high latency, increased energy consumption, significant communication overhead, and potential data privacy concerns. As a result, edge computing solutions, which facilitate localized processing and decision making, are gaining traction. Some segments of the Edge processing applications field are meaningful to be integrated into the image sensor, leading to the development of innovative image sensors that not only capture high-quality images but also perform on-chip processing to support low-latency and energy-efficient operations. One of the most promising approaches for meeting these requirements is three-dimensional (3D) integration. By vertically stacking multiple layers, 3D integration enables the heterogeneous combination of different technology nodes within a single compact form factor. This allows for the seamless integration of pixel arrays, readout circuits, memory, and dedicated processing logic to enhance system efficiency.

This paper presents J3DAI, a high efficiency and flexible artificial intelligence hardware accelerator designed for three-wafer stacked CMOS image sensor. J3DAI is partitioned into 3 dies, each serving a specific function: the top die houses the RGB pixel matrix, the middle die contains the analog-digital mixed signal design for pixel readout and processing, and the bottom die features the edge AI chip with a DNN accelerator optimized for multi-stacked edge AI CMOS image sensors. The DNN accelerator is designed to efficiently perform neural network tasks such as image classification and segmentation. Major innovations in data movement and MAC/cycle efficiency have been introduced to enhance the performance of the DNN system. This paper focuses on the digital system of J3DAI, highlighting its Performance-Power–Area (PPA) characteristics and showcasing advanced edge AI capabilities on a CMOS image sensor. To support hardware, a comprehensive software framework called Aidge [1] is used. Aidge enables the programming of both the host processor and the DNN accelerator, providing a seamless pipeline from model development to deployment on the target hardware. The framework supports post-training quantization, which is crucial for deploying models on resource-constrained hardware like J3DAI, significantly reducing memory footprint and computational complexity. The integration of advanced hardware and software into J3DAI represents a significant step forward in the development of efficient and powerful edge AI systems. This paper aims to provide a detailed exploration of the J3DAI architecture, its DNN accelerator, and the supporting software framework, demonstrating the potential of this innovative design in the field of edge AI.

This paper is organized as follows. Section II covers the related work and background, Section III details the J3DAI architecture, the design, and deployment tools. Section IV presents the experimental setup and results. Finally, Section V provides a conclusion.

## II. Related Work and Background

Three-dimensional (3D) integration has emerged as a key enabler for the development of advanced image sensors, overcoming the limitations of traditional two-dimensional (2D) designs. By stacking multiple layers vertically, 3D integration facilitates higher integration density, improved power efficiency, and reduced form factors, making it particularly beneficial for CMOS image sensors (CIS). This approach enables the heterogeneous integration of dedicated CMOS pixel technology with specialized processing circuits, enhancing both imaging quality and computational efficiency. Vivet *et al.* [17] presented significant advancements in 3D technologies and architectures for smart image sensors. These early contributions laid the groundwork for subsequent research in the field. Early research efforts on 3D-integrated image sensors focused primarily on optimizing image quality and sensor efficiency. Zhang *et al.* [18] introduced a feature extraction image sensor with a fill factor of 97% by separating photo-diodes and processing circuitry between different layers. This configuration maximizes light capture while ensuring efficient utilization of computational resources. Other pioneering efforts, such as those by Kiyoyama *et al.* [8], Ohara *et al.* [12], and Lee *et al.* [5], showcased block-parallel image sensing and processing techniques, significantly increasing throughput. Haruta *et al.* [6] integrated dynamic random access memory (DRAM) into the sensor stack to enable high frame rates and real-time electronic image stabilization. Similarly, Sukegawa *et al.* [15] used heterogeneous integration to fabricate sensor layers using 90nm technology while utilizing 65nm processing layers for enhanced performance. With the rise of Artificial Intelligence (AI) applications, recent research has focused on embedding Deep Neural Network (DNN) processing capabilities within 3D-stacked image sensors. One notable development is the "NeuroSensor" architecture [2], which integrates a Convolutional Neural Network (CNN) accelerator directly into the sensor stack. By enabling on-chip feature extraction and classification, this approach minimizes data transfer requirements and enhances energy efficiency.

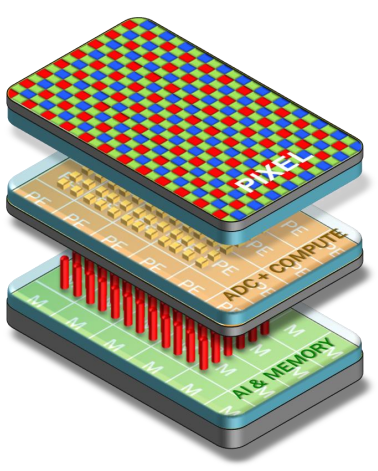

Fig. 1: Example of targeted functional 3-layer CIS device [11].

Beyond architectural innovations, advances in interconnection technologies are crucial to take advantage of the full potential of 3D integrated image sensors. Researchers in [16] explored 3D hybrid bonding techniques for multi-wafer stacking, utilizing high-density through-Silicon Vias (TSVs) with a 2 µm pitch to enable efficient inter-layer communication. Nicolas *et al.* [11] further demonstrated a three-layer test vehicle using Cu-Cu hybrid bonding, showcasing robust metal connexion and void-free interfaces for improved reliability and performance. An example of the targeted functional CMOS Image Sensor (CIS) is shown in Figure 1. These innovations collectively contribute to lower power consumption, enhanced computational accuracy, and increased processing speed. Rapid advancements in 3D-integrated image sensor technologies underscore the growing need for efficient and high-performance solutions in intelligent vision applications. By integrating advanced computational capabilities with optimized interconnect and noise reduction techniques, researchers are paving the way for a new generation of image sensors that cater to diverse applications, including autonomous systems, healthcare imaging, and next-generation IoT devices. This paper explores these developments in depth, identifying key challenges and future research directions to drive continued innovation in this field.

## III. J3DAI Architecture and AI Model Co-optimization

### *A. Architecture Description*

J3DAI is a 3-layer 3D-stacked edge AI CMOS image sensor architecture partitioned as follows:

The top die was designed to be an RGB pixel matrix with a resolution of 4096x3072 (12 million pixels), intended to occupy 16 mm² of silicon area (with dimensions of 4.7 mm in height and 3.4 mm in width, including the pads). To comply with the objective of defining an economically viable product, this device has been specified with the stringent constraint of being top die limited (in opposition to [4]): the budget allowed for middle and bottom layers are derived from this strong assumption. As a consequence, the budget for AI memory and processing power have been aligned on this constraint.

The second die includes the classical functionalities of a high quality 12-Mpixel image sensor, i.e. readout integrated circuits, signal processing, control and communication layers, and high speed interface (HSI) to transfer the full resolution image to a host processor when required, plus the capability to transfer sub-sampled images to the third layer, adapted with AI network capabilities. It is connected to the top pixel die with Cu-Cu hybrid bonding technology and is connected to the third die using High Density Through Silicon Vias (HD-TSV). The diameter of the HD-TSV is 1 µm and pitch is 2 µm as presented in [11], 6 mm² is for the analog readout integrated circuit and the rest of the area is dedicated to the Image Signal Processor (ISP), host processor (RISC-V 32b CPU) sub-system and part of L2 SRAM memory of the system.

The third die, still with the same dimensions, emphasizes the edge AI chip featuring the DNN accelerator. The DNN accelerator and memory components have been optimally sized to efficiently perform neural network tasks such as image classification and segmentation. The DNN system builds upon the approach proposed in [3] and has been previously applied in [9]. It has been optimized for multi-stacked edge AI CMOS image sensor. Furthermore, significant innovations have been introduced to improve data movement and MAC/cycle efficiency. These innovations are presented in III-B.

The J3DAI architecture partitioned into 3 dies is shown in Figure 2. The analog readout circuit and the pixel matrix are not the focus of this paper. This paper rather highlights the digital system and its PPA, while presenting advanced edge AI on the CMOS image sensor.

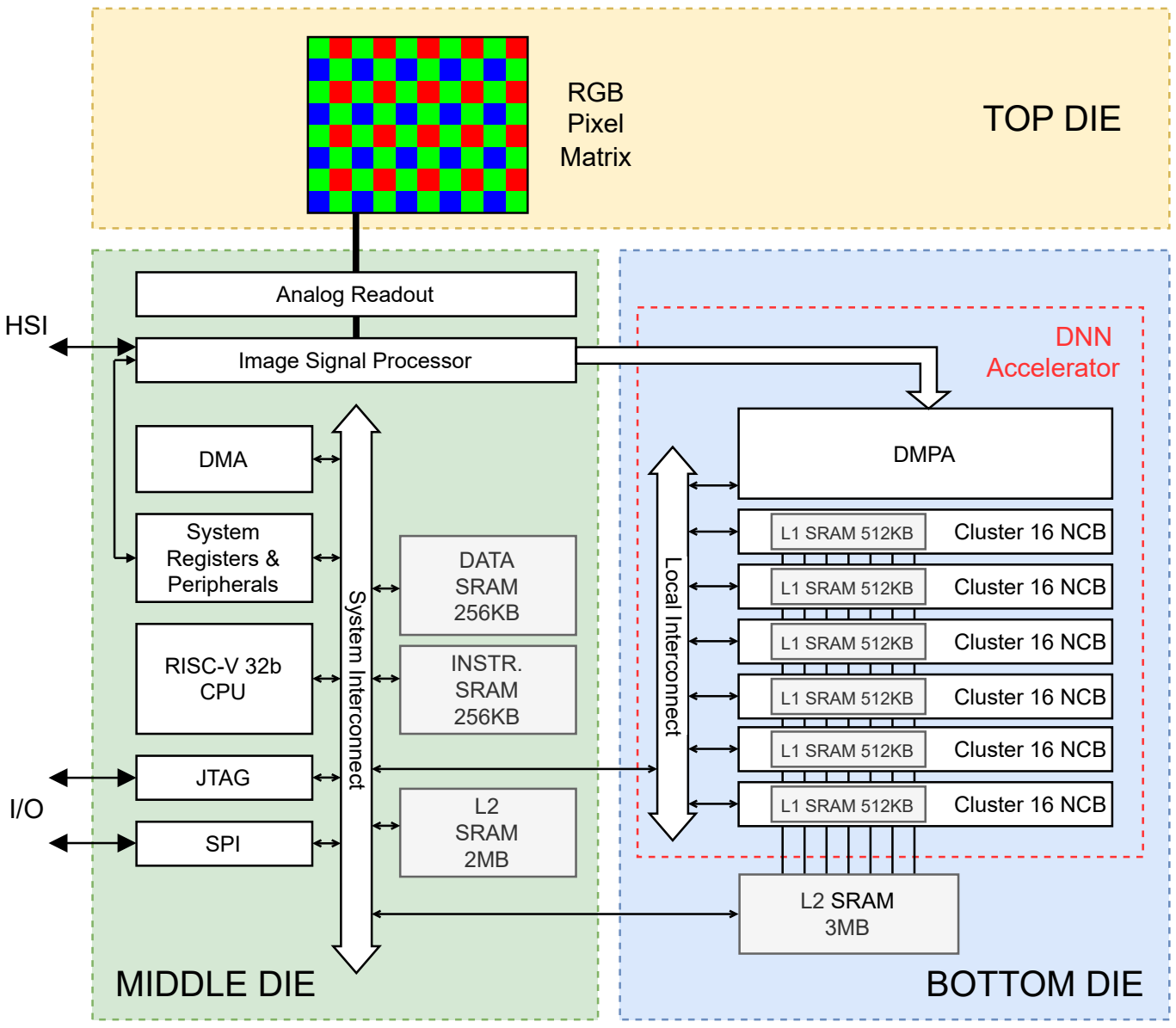

Fig. 2: J3DAI architecture and 3D partitioning

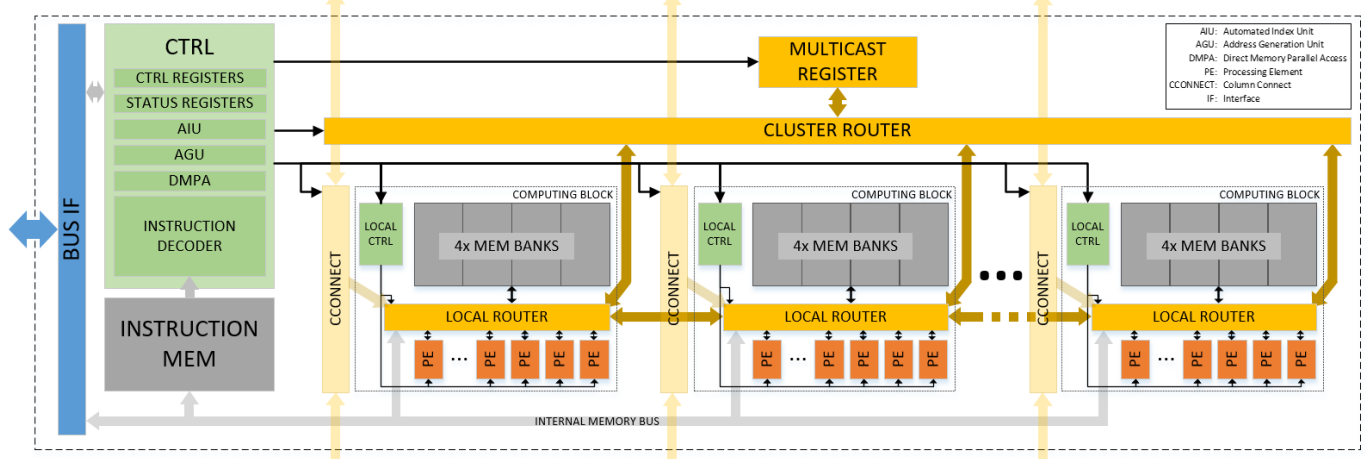

Fig. 3: Neural cluster architecture view

### B. DNN System Architecture

*1) System Integration:* The DNN system is mainly composed of a host processor, a DMA enabling efficient memory to memory transfer, a DNN accelerator, and a global memory (L2). All these components are connected through the system interconnect. The host is a RISC-V 32b CPU, with 256 KB of instruction memory and 256 KB of data memory. The DNN accelerator connects the host processor through the system interconnect. Defined as a near-memory architecture, all the internal memories are exposed to the whole system. Synchronization between the IP and the host is done through a set of registers and optional interrupt signals.

The architecture is scalable at multiple levels. The first scalability level is at the cluster level; every cluster can execute different programs or they can be synchronized by running a single program to increase the computing parallelism. The Neural Computing Block (NCB) scalability level allows one to adapt both computing power and transfer rate (cluster to cluster, cluster to/from global memory). From the host processor perspective, this memory is seen as a single block which can be used as a second level memory accessible through the system interconnect.

*2) Neural Cluster Architecture:* This programmable component is a SIMD architecture composed of neural computing blocks and cluster level routers, all controlled by a single unit. This module fetches and decodes instructions from the instruction memory. Control signals are broadcasted to all computing blocks. The local controller embeds all control and status registers accessible from the host processor through the system interconnect. The cluster architecture is shown in Figure 3.

The synergy of the Address Generation Unit (AGU), the Automatic Index Unit (AIU) and the multilevel data routers enable very efficient data movement. Thanks to the collection of these features, this neural accelerator is able to provide the right data to be computed at every cycle for a wide range of data distributions.

The Automatic Index Unit is based on configurable hardware loop and is able to automatedly drive the data routing selection. No additional instructions are required to configure the routing control. This reduces the program memory footprint and improves the number of operations per cycle.

The Direct Memory Parallel Access unit (DMPA) controls the Column Connect (CCONNECT) modules that are attached to the Computing Blocks or the global Memory Blocks. The global memory is tiled in multiple blocks that are arranged in symmetric columns according to the computing blocks of the neural clusters. These modules are capable of parallel vertical data movements in between clusters and global memory. The same control signals are broadcast to the CCONNECTs of a cluster or global memory. This feature improves data transfers that are required in the case of tiling when the memory size of the neural cluster is not sufficient, or also for copying parameters from the global memory to the accelerator memories. The DMPA enables the transfer of 1024 bits in a single clock cycle, or 1 MB in 1000 clock cycles. This performance is significantly superior to the limitations of DMA, which is constrained by the 64-bit width of the system interconnect bus in this case.

The association of the cluster router and the multicast register enables communications between computing blocks. This feature is able to multicast data from a given source to multiple destinations. Advanced routing features allow mixing of data coming from multiple sources. The multicast register has a direct path to one of the PE operands, giving access to the data in a single cycle.

*3) Neural Computing Block Architecture:* Neural computing block consists of a multi-banked static Random Access Memory (SRAM) connected to SIMD Processing Elements (PE).

The multi-bank SRAMs are composed of independent memories. No specific memory bank is dedicated to filter parameters or feature maps data. This flattened memory hierarchy allows to maximize the usable memory space and make it fully generic, no matter which application is executed. All internal memories can be accessed through the address generators of the cluster.

At the Neural Computing Block level, the local router module performs on-the-fly operations to transfer data between memories and PEs in a single cycle. It supports neighbor accesses, multi-cast transfers, and bit-shifting for data alignment between PEs and can introduce zeros or ones for padding operations. The multi-cast feature is helpful for sending the parameters to multiple PEs in a single cycle. It enables local data transfers of different precision formats without passing through the memories. Two local interconnects allow one to transfer 8-bit and the other 32-bit data. Computing Block routing modules are connected in a daisy chain. The programmer has dedicated instructions for configuring these connections.

The PE has been streamlined and optimized to efficiently perform the essential computations required by conventional convolutional neural networks. Each PE comprises a 9-bit multiplier, a 32-bit accumulator, an ALU (Arithmetic and Logic Unit) and a non-linear-operation unit based on an approximation of functions.

Taking into account the surface area on the last layer and the need to maintain the 4/3 aspect ratio of the image sensor, the best DNN accelerator configuration in terms of scalability was determined for the number of neural clusters and computing blocks. J3DAI features a DNN acceleratorwith 6 neural clusters of 16 computing blocks, each comprising 8 PEs. Thus, this configuration can output a maximum of 768 MAC (Multiply ACCumulate) operations per clock cycle.

### C. *Supporting Software*

*1) Aidge Framework:* To program the host processor and the DNN accelerator, a Software Design Kit (SDK) has been developed. This SDK is based on a specific Aidge export module. Aidge is a deep learning framework that supports the entire pipeline of model development, from definition and training to optimization and deployment on embedded platforms. The framework supports the entire pipeline of model development. In this work, the adapted strategy used PyTorch for model definition and training followed by export to ONNX [13]. The ONNX representation was then imported into Aidge, utilizing its optimization and deployment capabilities for the target hardware.

Aidge realizes post-training quantization, a critical step in deploying models on resource-constrained hardware like J3DAI DNN accelerator. Post-training quantization converts high-precision floating-point models (*e.g.* FP32) exported by PyTorch into low-precision fixed-point representations (*e.g.* INT8), significantly reducing memory footprint and computational complexity without requiring retraining. This process involves calibrating the model using a representative dataset to determine optimal scaling factors for weights and activations, ensuring minimal loss of precision during quantization.

Aidge generates code for different targets, including CPU, DSP, GPU with OpenMP, OpenCL, Cuda, cuDNN, FPGA, and various configurable DNN accelerators. A dedicated export module was developed to take advantage of the hardware features of J3DAI DNN accelerator, ensuring optimal performance and energy efficiency for quantized models. By combining the flexibility of PyTorch for model development and training with Aidge's robust quantization and deployment capabilities, the framework provides a comprehensive solution for implementing DNN-based applications on embedded platforms.

*2) DNN accelerator Export:* The export starts by analyzing the application parameters and the hardware configuration. A dedicated solver explores multiple mapping solutions to find the optimal data memory placement. According to this placement, Processing Elements (PEs) are assigned to define a computing process. Then, the computation is mapped to processes that operate in parallel. The solver minimizes the need for data movement during computation to achieve the best operation per cycle rate. It checks if the data fit in memory and generates metrics like computing resource usage. The scheduling optimization solver looks for the best way to mask parameter loading. At every execution step, it verifies if an additional memory bank is available and explores multiple schedules to minimize execution time. Based on the mapping, the export sets the configuration of each advanced feature like automatic indexes and multidimensional address generators.

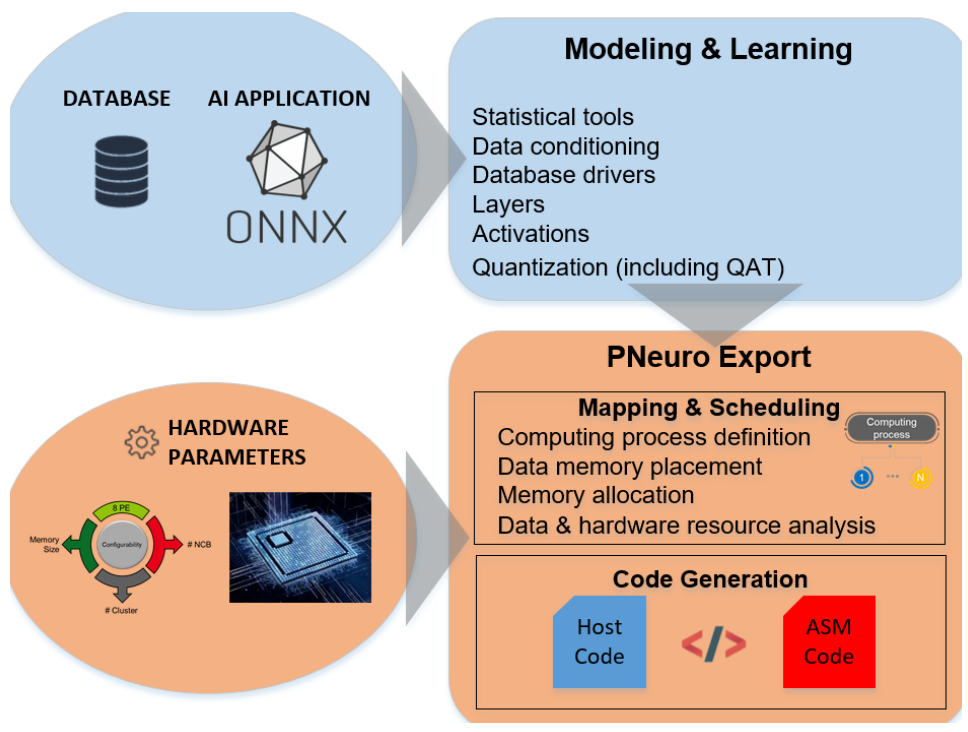

Fig. 4: DNN accelerator export

After the mapping metrics have been defined, the export automatically generates the host and assembly codes. The host code is composed of the main source code that will call the function of each layer with an execution routine. Figure 4 provides an overview of the export.

## IV. Experiment and Results

### A. *Implementation*

A J3DAI implementation has been realized in 28nm FDSOI technology. The target frequency was 200 MHz. The synthesis of the bottom and middle layers were performed using Synopsys Design Compiler, and the placement and routing were realized with Cadence Innovus. In the middle die, 6 $mm^2$ have been allocated for the analog readout controlling the pixel matrix. The floorplans of the middle and bottom dies are given by Figure 5.

The maximum amount of L2 memory has been placed in the bottom layer, totaling 3 MB. An additional 2 MB has been placed on the middle layer. Therefore, the total embedded L2 memory is 5 MB, which enables the execution of several networks that require multiple MBs to store parameters.

The middle and bottom dies are connected through 3K through-silicon vias (TSV). 2048 of these vias are used to establish data connections between L2 memory partitions of the bottom and middle layers. Indeed, the design integrates 16 memory blocks, each 64-bit wide, which requires 1024 bits for data transfer from the middle layer to the bottom layer and 1024 bits in the opposite direction.

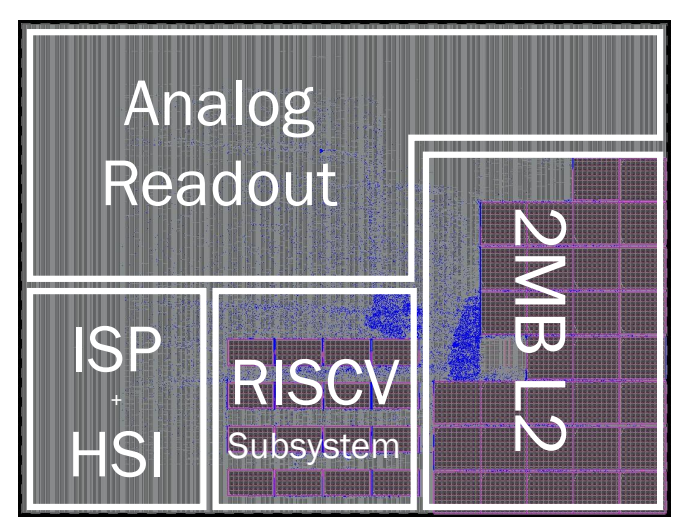

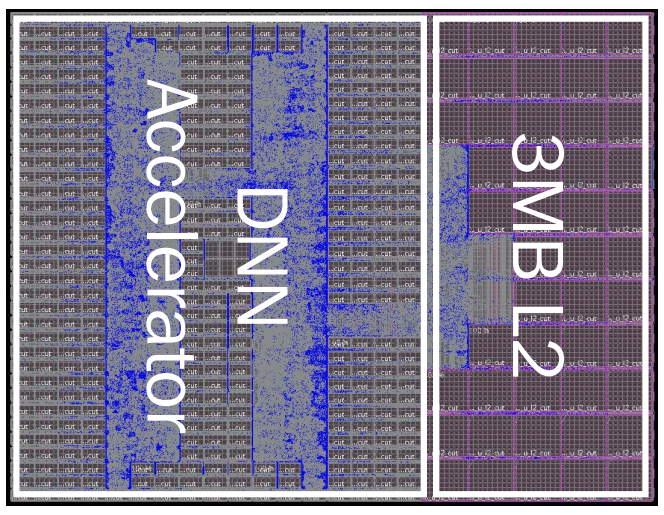

Fig. 5: J3DAI Middle (a) and Bottom (b) die floorplans

TABLE I: Keys performances metrics of selected models

| Model | MobileNetV1 | MobileNetV2 | Segmentation |
|---|---|---|---|
| **MMACs** | 557 | 289 | 877 |
| **Image Input** | 256x192 | 256x192 | 512x384 |
| **Accuracy** | 62% (Top-1) | 66% (Top-1) | 36.4 (mIoU) |
| **Latency @200MHz** | 4.96 ms | 4.04 ms | 7.43 ms |
| **Power @30FPS** | 47.6 mW | 30.5 mW | 63.8 mW |
| **Power @200FPS** | 291.2 mW | 186.7 mW | - |
| **Power efficiency** | 0.77 TOPs/W | 0.62 TOPs/W | 0.82 TOPs/W |
| **MAC/Cycle efficiency** | 76.8% | 46.6% | 76.5% |

### *B. Model Architecture*

To demonstrate the capabilities of J3DAI DNN accelerator, we used two distinct DNN models: a lightweight feature extraction network and a more complex segmentation network. These models were chosen to showcase the hardware versatility in handling both simple and computationally intensive tasks.

*1) Feature Extraction with MobileNetV1 and V2:* For feature extraction, we use MobileNetV1 [7] and MobileNetV2 [14], which are lightweight architectures designed to optimize inferences on resource-constrained devices. These models consist of depthwise separable convolutions and linear bottlenecks, significantly reducing the number of parameters and computational complexity. The use of ReLU activation functions ensures compatibility with post-training quantization, enabling efficient deployment in J3DAI DNN accelerator. We used ImageNet pretrained quantized weights (with uint8 precision) for both MobileNetV1 and V2, ensuring strong baseline performance for classification tasks. Their simplicity and efficiency make them ideal for demonstrating J3DAI ability to handle low-latency, high-throughput tasks.

To cope with the 4/3 sensor form factor, the input image size is scaled to 256x192 pixels. For this configuration, the number of MAC operations for MobileNetV1 is 557 million, which is comparable to the standard format that uses images of size 224x224 (569 million MACs). Similarly, for MobileNetV2, the number of MACs is 289 million, compared to 300 million in the standard format using images of size 224x224 pixels. The MAC/cycle efficiency ranges from 46.6% to 76.8%. The MobileNetV2 network contains branching structures, which introduce additional data movement, increasing latency, and reduces MAC/cycle efficiency. In contrast, MobileNetV1 has a simpler convolution-based architecture, resulting in higher MAC/cycle efficiency. However, it requires more computations (557 MMACs vs. 289 MMACs) to achieve a similar level of accuracy.

It is important to note that both networks are tested on a scaled input image size (256x192) for which they were not originally trained. As a result, their accuracy is lower compared to state-of-the-art models for the same precision. However, these models can be fine-tuned and adjusted for specific applications, making them highly adaptable for targeted use cases.

*2) Adapted Segmentation Network:* This more complex model demonstrates the ability of J3DAI to handle demanding workloads, such as pixel-level prediction tasks. For this purpose, we designed and trained a segmentation network based on the Feature Pyramid Network (FPN) architecture. The model was trained on the Cityscapes dataset, a widely used benchmark for urban scene understanding and semantic segmentation tasks. To fit the memory and computational constraints of J3DAI, we adapted the FPN backbone and the segmentation head by reducing the depth of the convolutional layers. Furthemore, we scaled the width multiplier ($\alpha$) of MobileNetV1 to 0.5, which controls the number of channels in the feature extraction layers. These modifications ensured that the model needs 877 MMACs and could operate efficiently within the hardware's memory and computational limits while maintaining competitive accuracy, 36.4 mIoU on the cityscapes. To take advantage of transfer learning, the segmentation model was initialized with ImageNet pretrained weights for MobileNetV1 and fine-tuned on the Cityscapes dataset. This approach allowed the model to benefit from robust feature representations while adapting to the specific requirements of urban scene segmentation.

All models were initially defined and trained in PyTorch. After training, the models were exported to the ONNX format and imported into Aidge for post-training quantization and hardware-specific optimizations. The networks employ ReLU activation functions throughout, ensuring compatibility with post-training quantization methods. The quantized models were then deployed on J3DAI, demonstrating the platform's capability to handle advanced computer vision tasks with efficient resource utilization.

### *C. Results and comparison*

For power consumption estimation, all experiments are conducted using post-place and route netlists. Both sub-systems operate at 0.85V. Siemens Questasim is used to perform the gate level simulation and generate VCDs. Synopsys PrimePower tool is used for power analysis.

The Table I provides a comprehensive overview of the performance characteristics of MobileNetV1, MobileNetV2, and Segmentation models in several key performance metrics.

The Table II provides a comprehensive comparison of three advanced imagers: SONY ISSCC 2021 [4], SONY IEDM 2024 [10], and J3DAI. The table is organized to offer a side-by-side comparison of key technical specifications and performance metrics for each imager.

Regarding the fabrication process, [4] employs a two-layer fabrication process with a top chip at 65nm and a bottom chip at 22nm but is suffering from a significant bottom die limitation

TABLE II: Comparison table with prior works

| | **SONY ISSCC'2021 [4]** | **SONY IEDM'2024 [10]** | **This Work [J3DAI]** |
|---|---|---|---|
| **Chip characteristics** | | | |
| **Fabrication Process (Top / Middle / Bottom)** | 65nm / n.a. / 22nm | 65nm / 40nm / 22nm | 40nm / 28nm / 28nm |
| **Chip size** | 7.558 mm (H) x 8.206 mm (V)<br>124 mm$^2$ | 11.2 mm (H) x 7.8 mm (V)<br>262 mm$^2$ | 4.698 mm (H) x 3.438 mm (V)<br>48 mm$^2$ |
| **DNN + Internal memory size** | 31 mm$^2$ (estimated 50% bottom chip) | 87 mm$^2$ | 16 mm$^2$ |
| **Number of effective pixels** | 4056 (H) x 3040 (V) | 8784 (H) x 6096 (V) | 4096 (H) x 3072 (V) |
| **Pixel size** | 1.55 µm (H) x 1.55 µm (V) | 1.12 µm (H) x 1.12 µm (V) | 1 µm pixel pitch |
| **Logic supply voltage** | 0.8V | 0.8V, 1.1V | 0.85V |
| **DNN System** | | | |
| **Processor Clock** | 262.5 MHz | 219.6 MHz | 200 MHz |
| **Number of MACs** | 2304 MACs | 1024 MACs | 768 MACs |
| **MAC processing efficiency* [%]** | 13.4 | 59.9 | 46.6 |
| **Power Consumption* [mW]@200fps** | 122.5 | 90.4 | 186.7 |
| **Processing time* [ms] @ 262.5 MHz** | 3.70 | 1.87 | 3.01 |
| **Power efficiency*** | 0.98 TOPS/W | 1.33 TOPS/W | 0.62 TOPS/W |
| **Energy efficiency per unit area* (GOPS/W/mm$^2$)** | 7.9 | 5.1 | 12.9 |

* Remarks: MobileNetV2

(appx 30%). In contrast, both [10] and J3DAI utilize three-layer architectures. [10] features a top chip in 65nm, a middle chip in 40nm, and a bottom chip in 22nm. The J3DAI has a top chip in 40nm, a middle chip in 28nm, and a bottom chip in 28nm. [4] has a chip size of 7.558 mm (H) x 8.206 mm (V), resulting in an area of 124 mm$^2$. [10] is larger, with a chip size of 11.2 mm (H) x 7.8 mm (V), covering an area of 262 mm$^2$. The J3DAI is the most compact, with a chip size of 4.698 mm (H) x 3.438 mm (V), occupying an area of 48 mm$^2$. Figure 6 shows a graphical representation of the three sensors at scale.

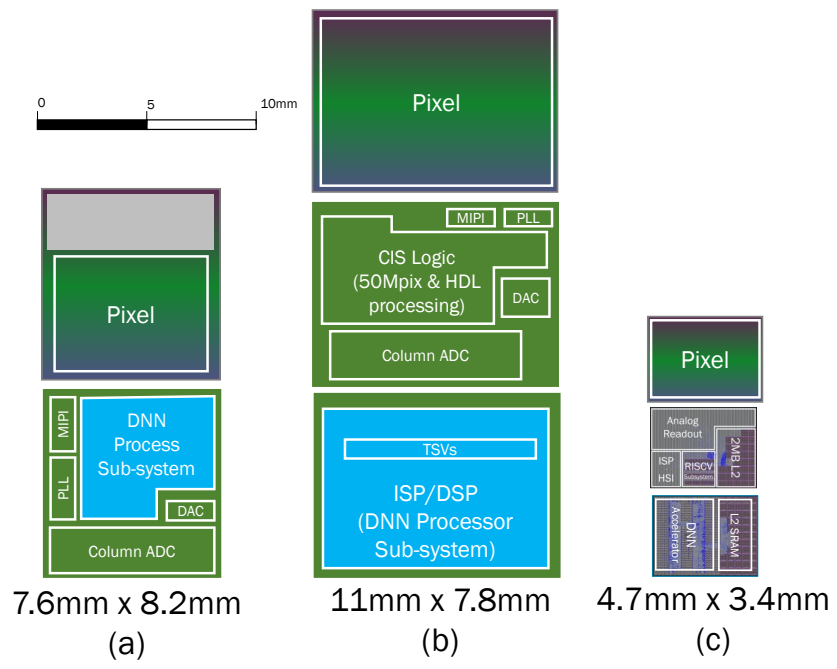

Fig. 6: Chip-size comparisons of (a) 2-layer stacked [4], (b) 3-layer stacked [10], (c) 3-layer stacked [This work]

J3DAI, with only 768 MACs, is the most integrated, occupying only 16 mm$^2$, compared to 31 mm$^2$ for [4] (2304 MACs) and 87 mm$^2$ for [10] (1024 MACs). J3DAI achieves a MAC efficiency per cycle of 46.6%, outperforming [4] at 13.4% but trailing [10] at 59.9%. For J3DAI, higher efficiencies have been observed in networks with different topologies, such as MobileNetV1, which reaches 76.8%.

Although J3DAI has the highest power consumption at 186.7 mW, compared to 122.5 mW for [4] and 90.4 mW for [10], its high level of integration allows it to achieve the best energy efficiency per unit area, reaching 12.9 GOPS/W/mm$^2$, versus 7.9 GOPS/W/mm$^2$ for [4] and 5.1 GOPS/W/mm$^2$ for [10].

## V. Conclusion

This paper presents J3DAI, a high-efficiency and flexible artificial intelligence hardware accelerator designed for a three-wafer stacked CMOS image sensor. J3DAI is partitioned into three dies, each serving a specific function: the top die houses the RGB pixel matrix, the middle die contains the classical functions of a high quality image sensor, and the bottom die features the edge AI chip with a DNN accelerator optimized for multi-stacked edge AI CMOS image sensors.

The DNN accelerator is designed to efficiently perform neural network tasks such as image classification and segmentation. Major innovations in data movement and MAC/cycle efficiency have been introduced to enhance the performance of the DNN system. This paper focused on the digital system of J3DAI, highlighting its performance-power–area (PPA) characteristics and showcasing advanced edge AI capabilities on a CMOS image sensor.

To support hardware, we used the Aidge comprehensive software framework, which enables programming of both the host processor and the DNN accelerator. Aidge supports post-training quantization, significantly reducing memory footprint and computational complexity, making it crucial for deploying models on resource-constrained hardware like J3DAI.

The integration of advanced hardware and software into J3DAI represents a significant step forward in the development of efficient and powerful edge AI systems. Our experimental results demonstrate the versatility and efficiency of J3DAI, showcasing its potential in handling both simple and computationally intensive tasks.

Future work will focus on further optimizing the architecture and exploring new applications to fully leverage the capabilities of J3DAI. As edge AI continues to grow in importance, innovations like J3DAI will play a crucial role in enabling real-time, low-latency, and energy-efficient edge AI processing.

## Acknowledgments

This work was supported by the French National Research Agency (ANR) through the Investissement d'avenir" (investments for the future) programs: ANR 10-AIRT- 005 (IRTNANO-ELEC).]